# Gender differences in time perception and its relation with academic performance: non-linear dynamics in the formation of cognitive systems

**Klaus Jaffe [2], Guillermo Mascitti [1], Daniella Seguias [2]**

[1] Istituto Cantonale di Economia e Commercio, Bellinzona, Switzerland

[2] Universidad Simón Bolívar, Caracas, Venezuela

*Correspondence: Klaus Jaffe, Universidad Simón Bolívar, Apartado 89000 Caracas 1080, Venezuela.*
*E-mail: kjaffe@usb.ve*

# Abstract

**Introduction**: Non-linear dynamics is probably much more common in the epigenetic dynamics of living beings than hitherto recognized. Here we report a case of global bifurcation triggered by gender that affects higher cognitive functions in humans.

**Method**: We report a cross-cultural study showing deviations in time perception, as assessed by estimating the duration of brief sounds, according to their durations and to the gender of the perceiver.

**Results**: Duration of sounds lasting less than 10 s were on average overestimated, whereas those lasting longer were underestimated; estimates of sounds shorter than 1 s were extremely inaccurate. Females consistently gave longer estimates than males. Accuracy in time estimation was correlated to academic performance in disciplines requiring mathematical or scientific skills in male, but not in female students. This difference in correlation however had nothing to do with overall skills in mathematics. Both sexes scored similarly in scientific and technical disciplines, but females had higher grades than males in languages and lower ones in physical education.

**Conclusion**: Our results confirm existing evidence for gender differences in cognitive processing, hinting to the existence of different "mathematical intelligences" with different non-linear relationships between natural or biological mathematical intuition and time perception.

# Resumen

**Introducción**: La dinámica no lineal es posiblemente mucho más común de lo que se reconoce actualmente. Aquí reportamos un caso de bifurcación global en la historia epigenetica del individuo modulado por el sexo del individuo y que tiene repercusiones cognitivas superiores en humanos.

**Método**: Reportamos un estudio multi-cultural sobre desviaciones en la percepción del tiempo, medido como estimados de la duración de un sonido breve, según el tiempo de duración y el sexo del estimador

**Resultados**: El tiempo de los sonidos con duración menor a 10 s fueron mayormente sobreestimados, mientras que los sonidos de duración mayor fueron subestimados; estimaciones de tiempo de sonidos de menos de 1 s fueron muy inexactas. Las hembras estimaron las duraciones de los sonidos consistentemente como mas largas que las estimaciones hechas por varones. La precisión de los estimados de tiempo correlacionó con las notas académicas relacionadas con las matemáticas o las ciencias naturales en varones más no en las estudiantes femeninas. Esta correlación sin embargo resultó ser independiente de las habilidades matemáticas absolutas del individuo. Amos sexos tenían notas similares en disciplinas técnicas y científicas pero las hembras tenían notas superiores a los varones en lenguaje y notas menores en educación física.

**Conclusiónes**: Los resultados confirman las evidencias existentes sobre las diferencias cognitivas entre los sexos, sugiriendo la existencia de "inteligencias matemáticas" diferentes con una relación no lineal entre la intuición matemática biológica o natural y la percepción del tiempo.

**Palabras clave:** genero, sexo, tiempo, percepción, matemática, ciencia, aprendizaje, cognitivo

# Introduction

Catastrophe theory, which originated with the work of the French mathematician René Thom (1994) postulates that small changes in critical parameters of a nonlinear system can drive the system to attractors or equilibrium states evidences by a sudden changes of the behavior of the system. Often such changes occur when the system dynamics offers two paths to follow. Thus we speak if a bifurcation which occurs when a small smooth change made to the parameter values (the bifurcation parameters) of a system causes a sudden 'qualitative' or topological change in its behavior. When larger invariant sets of the system 'collide' with each other, or with equilibria of the system we call the phenomena a global bifurcations. These type of bifurcations cannot be detected purely by a stability analysis of the equilibria. Catastrophic dynamics in non-linear systems are ubiquitous and well described in physics. Their existence in psychological systems has been suggested and some miscellaneous cases have been described (Petitot, 1989, Van der Maas and Molenaar 1992, Poston and Stewart 2007, for example). Yet their relevance for understanding cognitive processes remains to be demonstrated as few psychological phenomena been recognized as being governed by such non-linear dynamics. Here we present a bifurcation phenomenon, triggered by gender, which is best understood under the light of René Thom´s catastrophe theory as applied to the genesis of cognition.

For many different human attributes such as height, weight, propensity for criminality, overall IQ, mathematical ability, and scientific ability, there is clear evidence for gender differences, not only in the mean value of these attributes, but also in their standard deviation and variability. In some cases, we have evidence that the different hormones males and females are exposed to during fetal development may explain this difference (Kimura, 2002; Kempel et al., 2005). From a biological perspective, sex differences in behavior are an adaptive resource and as such should yield a discernible advantage for each sex (Jacobs et al., 1990). In the domain of cognitive skills, the best documented of these differences are those in spatial perception, which promote sex asymmetries in a vast range of abilities. More specifically, females perform better in tasks involving production and comprehension of language, fine motor skills, and perceptual speed, whereas males have a higher capacity for managing spatial relationships and carrying out visuospatial transformations (Linn & Petersen, 1985; Nagae, 1985).

These differences seem to favour the professional success of men in the hierarchical ladder of vertically integrated organizations, which require taking fast decisions with little input from peers or from the group. Here, abilities to abstract complex scenarios will be advantageous. These same differences favour women in institutions with a horizontal organizational structure, rewarding typically female skills such as attention to detail and to simultaneously occurring events (Briceño & Jaffe, 1997). Thus, differences in basic cognitive processes such as space perception appear to affect personal and social success, and foster composite gender differences in sociocultural expressions.

Space and time are sides of the same coin. Evidence exists that the two sexes differ not only in spatial perception, but also in elementary time processing abilities such as duration estimation (see reviews in Rammsayer & Lustnauer, 1989; Block & Zakay, 1997; Block et al., 2000; Glickson, 2001; Grondin, 2001; Walsh, 2003). More and more evidence accumulates showing that these differences have a neurophysiological basis (Harrington et al., 1998). The main finding is that females make larger time estimates than males. One possible explanation (Block et al., 2000) is that females pay more attention to time, thereby experiencing dilated subjective durations—something akin to what happens when we sit in the dentist's waiting room. For many researchers (Xie & Shauman, 2003) the timing issue is fundamental for psychology, as efficient performances depend partly on the capacity to make time -related adjustments. Namely, the question arises as to how one might keep track of time not only to minimize dispersion but also to remain close to the target duration (Grondin, 2001).

Alterations in time perception have been found in patients with language and reading disorders, such as children with dyslexia (Nicolson et al., 1995) or Attention-Deficit Hyperactivity Disorder (Smith et al., 2002), indicating that deficits in temporal perception of very brief intervals may interfere with higher-level functions such as language and reading skills. Both clinical and neurological studies suggest that the same deficits may also affect negatively motor timing and, as a consequence, motor skills (Ivry & Keele, 1989; Keele et al., 1985). This evidence is consistent with the possibility that, as in the case of space perception, small and apparently trivial individual differences in time perception underlie salient differences in the complex skills subserving more complex personality traits (Rammsayer & Rammstedt, 2000) and academic and professional success. An important limitation if we want to explore further these ideas is the heterogeneity of research approaches and thus of results in the literature.  For example Block et al., 2000 conclude from their meta-analysis that prospective judgment of time duration showed no overall sex effect whereas retrospective judgments showed a larger subject to object duration for females than for males. Later publications, such as Espinosa-Fernández et al., 2003, for example using an empty interval production task, report a greater underproduction of longer intervals (1 and 5 min) for women. These are only two examples of large number of papers tackling gender differences in time perception, each using a particular class of subjects and with distinct methodology. Thus it is unclear how results showing gender differences in time perception is due to cultural characteristics, age, methodology used or the analytical approach of the researcher.

Here we want to explore the potency of ethological approaches in prospective judgment of time using the duration of a sound presented to students of similar age in widely different cultures. Once detected, we correlate these differences with measurable cognitive performance of the individuals.

**Methods**

We tested 228 male and 273 female students at the Istituto Cantonale di Economia e Commercio, Bellinzona (Ticino), Switzerland, for their accuracy in estimating the duration of a sound produced by an electric bell. Subjects ranged in age between 15 and 19 years (mean age = 16.5 years). In groups of 15 to 25, we explained them the experimental procedure asking them to estimate the duration of the sound of an electric whistle. Then they were presented with a sequence of three sounds. The three sounds lasted respectively 5, 12, and 20 seconds, and where separated by a 20-second interval during which participants wrote down their estimate of the duration of the preceding sound. A sample of 232 was tested separately for their accuracy in time estimation of sounds lasting 0.5, 1, 3, 5 and 12 s, presented in random order, using the method described above.

Final grades for each student in Italian, English, German, Mathematics, Natural Sciences, Economics, and Physical Education, on a scale of 1 to 6 (worst to best) were also collected, together with the mean of all the grades obtained during the previous year and the total number of negative ('fail') grades. Seven subjects were excluded from the analysis because they made errors larger than 200% in time estimates.

Another test was performed with a group of 134 males and 380 females aged between 20 and 30 years (mean age = 22.9 years), from 9 different careers or faculties at Universidad Andres Bello, Caracas, in 2002. These were 1: Education (9 males; 106 females), 2: International relations (9; 40), 3:

Philosophy (8; 14), 4: Sociology (34; 64), 5: Accounting (19; 29), 6: Psychology (1; 20), 7: Economy (11; 15), 8: Social communication (13; 70), 9: Engineering (30 males; 22 females). A further division was made between careers with little mathematical requirements (Humanities) and those with mathematical requirements (Calculus based). In the first group we included students from careers 1, 2, and 4; whereas in the second group we included students from careers 5, 7 and 9. Subjects were given a Spanish translation of 40 items from a questionnaire (Zimbardo & Boyd, 1999) which explores personality differences based on subjective appreciation of temporal projections. In addition they were asked to estimate, by writing it down, the duration of a 3 s sound produced by an electronic whistle. Data on the students' grades (academic index) were provided by the university administration.

Accuracy in time assessment was estimated using two types of errors: The *scalar error* was the time interval estimated by the subject minus the time interval measured with a chronometer. The *absolute error* was the absolute value of the scalar error expressed as a percentage of the real duration of the sound.

**Results**

The accuracy of time perception depended on the length of the sound. The scalar error of time estimates is presented in Figure 1. Time estimates overshoot the real duration of the sound for short time intervals (less than 10 s) and undershoot them for longer ones (more than 10 s).

The longer the sound, the larger was the variance of the estimates. For example, the standard error of the percent deviation of the estimates from the actual duration of the sounds was 6.2, 3.3, 2.1, 1.5, 1.0, 0.01 for 0.5, 1, 3, 5, 12 and 20 s respectively, showing that time estimation became progressively more difficult as the stimulus shortened.

Females made consistently longer time estimations than males, independently from the duration of the sound, $(F(3,489)=4.7, p=0.003)$. The sum of the absolute errors for 3, 5, and 12 s for our full sample was roughly the same for males and females; but the scalar error differed between the sexes (Table 1). The absolute error correlated with the scalar error for females, $r(267)=0.33, p<0.001$, but not for males, $r(226)=-0.007, p=0.91$.

Academic performance also differed between the sexes, as can be seen from Table 1. Females had higher grades in Languages, but not in Scientific or technical disciplines; males outperformed females in Physical Education.

Absolute error in estimating time correlated significantly and negatively with grades in mathematics for males, r(223)=-0.18, p=0.007, but not significantly for females, r(264)=-0.06, p=0.354. Scalar error correlated negatively with grades in Italian for males, r(222)=-0.18, p=0.007, but not for females, r(265)=-0.05, p=0.42. In females, absolute errors did not correlate with grades in any discipline, as shown in Figure 2. This figure shows that male students making larger absolute errors had also lower grades in mathematics. The accuracy in time perception among males measured by the absolute error, correlated slightly with grades in English, r(224)=0.15, p=0.03.

Interestingly, very short (0.5 s) and very long sounds (20 s) correlated differently with academic success. Absolute errors in the estimation of very short sounds correlated inversely with grades in natural sciences (r(74)=-0.31, p<0.001) and economics (r(74)=0.31, p=0.008) for males, but not for females (p>0.05 for all disciplines). Absolute errors in the estimation of very long sounds correlated negatively with grades in mathematics in males (r(223)=-0.14, p=0.3) but not in females (r(264)=-0.04, p=0.51). That is, in long time intervals, where counting helps, there was a correlation between grades in mathematics and accuracy in time perception among males. In short time intervals this correlation disappeared and a correlation with natural sciences appeared. If the average grades of both mathematics and natural sciences are pooled and correlated with scalar error, the only significant (negative) correlation that emerges is among males, r(226)=-0.17, p=0.01. No correlation emerges among females (p>0.3).

Age and culture might affect these results. Thus, we include results of a different study using a separate sample of older students at a University in Caracas. Here, again, females produced longer estimates for the 3 s sound than males (t = 2.44, p = 0.014, df= 512), and obtained higher grades in their academic index than males (t=3.6, p= 0.0004, df=512). Scalar errors depended more on gender than on the faculty the student was enrolled (Figure 3). That is, in all cases, female students reported larger errors than males. Interestingly, the correlation between cognitive skills and errors in time perception, again, was measurable in males but not in females. Male students in faculties which require few mathematical skills, such as Education, Sociology, and International Relations tended to overestimate time, i.e. made large scalar errors, whereas male students in Engineering, Economy and Accounting tended to underestimate it. Females' errors did not differ statistically between the two groups of academic careers (Table 2).

As to the personality questionnaire, again results for females differed from those for males (Seguias, 2002). The most divergent results was that errors in time estimation correlated positively with a question measuring negative past ("It's hard for me to forget unpleasant images of my youth") in males (Spearman R = 0.22, p = 0.014, n = 129), but not in females (Spearman R = 0.0001, p = 0.99, n = 358), and with one question measuring positive past ("I enjoy stories about how things used to be in the 'good old times") in females (Spearman R = 0.15, p = 0.004, n = 363) but not in males (Spearman R = 0.10, p = 0.26, n = 128), confirming that our method was sensitive enough to detect known correlations between time estimation are personality traits in the two sexes.

**Discussion**

Our study confirmed the existence of gender differences in the prospective perception of time, where females tend to give longer estimates than males (Block et al., 2000). This difference remained constant over the whole range of durations we used that is from half a second to twenty seconds. On average, subjects overestimated intervals shorter than 10 seconds, and underestimated intervals longer than 10 seconds. These results are consistent with several findings summarized by Grondin (2001) and those reported by Collaer and Hill (2006), who reported that visuospatial performance increased with mathematical competence. The fact that for very short intervals, the variance of time estimates increased considerably, suggests that humans are not adapted to consciously estimate time intervals shorter than 1 second.

It has been suggested (Block et al., 2000) that the estimation of intervals depends on the attention deployed on passing time, and that females give larger estimates than males because they focus attention on time more, hence accumulating subjective temporal units at a faster rate. This, however, fails to explain the correlation between time estimation accuracy and mathematical skills in males but not in females. Nor does it explain the differences in time perception between males enrolled in faculties requiring or not requiring mathematical skills, and the lack of difference in females. Our results suggest that cognitive processing of time is different for the two sexes. Existing theories, including simple models trying to integrate all psychological mechanisms of time perception, such as those reviewed by Grondin (2001) and others mentioned in the introduction, cannot explain the results presented here.

Our results touch complex and poorly understood cognitive processes (Fink & Neubauer, 2001; Buhusi & Meck, 2005; Shaw et al., 2006). For example, behavioral and brain-imaging experiments (Dehaene et al., 1999) hint that mathematical intuition depends on linguistic competence or visuo-spatial representations. Exact arithmetic is acquired in a language-specific format and recruits networks involved in word-association processes. In contrast, approximate arithmetic shows language independence, relies on a sense of numerical magnitudes, and recruits bilateral areas of the parietal lobes involved in visuospatial processing. Our results showed that the sexes differ in the relationship between time assessment and mathematical skills suggesting gender differences in basic neurophysiological processing of cognitive tasks. These effects then might have further consequences on more derived aspects of individual and social behavior such as personality (Rammsayer & Rammstedt, 2000; Rammsayer, 1997)

We feel that the present state of the art does not allow for a meaningful detailed interpretation of our results. They can be only understood in the context of non-linear relationships between perception and cognition where biological based physiological processes have an additional non linear effect on this relationship. This study allows for some general conclusions:

1- Time perception is reasonably accurate within a window of durations ranging from a few seconds to a few minutes;

2- Genders differ consistently in time estimates, with females, on average, giving longer estimates;

3- We add to the evidence that time perception accuracy correlates with different personality traits, as also suggested by Rammsayer and Rammstedt (2000);

4- Skills in time perception are related to cognitive performance in ways that differ between the sexes;

5- The fact that our main findings were evidenced with subjects of different ages and cultural environments support the idea that gender differences in time perception are biologically, rather than culturally based (see Kimura, 1999).

The phenomenon here described can be understood if we accept that non-linear phenomena, similar to those described by René Thom in his application of catastrophe theory to morphogenesis, apply to the construction of cognitions. Although the details of the dynamics casing the phenomena described are still to be uncovered, the results are compatible with a vision in which physiological equilibria determined by gender affect basic processes underlying perception (time perception for example), which in turn affect the development of cognitive systems. It seems reasonable to suggest that this kind of non-linear dynamics guides many more common behavioral systems than currently recognized.

**References**

Block, R.A.& Zakay, D. (1997). Prospective and retrospective duration judgments: A meta-analytic review. *Psychonomic Bulletin & Review* **4** 184-197

Block, R.A., Hancock, P.A. & Zakay, D. (2000). Sex differences in duration judgments: A meta-analytic review. Memory and Cognition, 28: 1333-1346

Buhusi, C.V. & Meck, W.H. (2005) "What makes us tick? Functional and neural mechanisms of interval timing" *Nature Reviews* *6* 755-765

Bressan, P. (2002). The connection between random sequences, everyday coincidences, and belief in the paranormal. Applied Cognitive Psychology, 16: 17-34.

Briceño, A. & Jaffe, K. (1997). Sex differences in occupational performance: a biological perspective. Social Biology, 44: 198-204.

Buss, D. (1989). Sex differences in human mate preferences: Evolutionary hypotheses tested in 37 cultures. Behav. Brain Sci. 12:1-49.

Collaer, M.L. & Hill, E.M. (2006). Large sex difference in adolescents on a timed line judgment task: Attentional contributors and task relationship to mathematics. Perception, published online 8 March 2006

Dehaene, S., Spelke, E., Pinel, P., Stanescu, R. & Tsivkin, S. (1999). Sources of Mathematical Thinking: Behavioral and Brain-Imaging Evidence. Science 284: 970-974.

Espinosa-Fernández, L., Miró, E., Cano, M. & Buela-Casai, G. (2003). Age-related changes and gender differences in time estimation. Acta psychologica, 112: 221-232.

Fink, A. & Neubauer, C. (2001) Speed of information processing, psychometric intelligence, and time estimation as an index of cognitive load. *Personality and Individual Differences* **30** 1009-1021

Geary, D.C. (1996). Sexual selection and sex differences in mathematical abilities. Behavioral and Brain Sciences 19: 229-284.

Gilger, J.W., & Ho, H.Z. (1989). Gender differences in adult spatial information processing: Their relationship to pubertal timing, adolescent activities, sex-typing of Personality. Cognit. Develop. 4:197-214.

Glickson, J. (2001). Temporal cognition and the phenomenology of time: A multiplicative function for apparent duration. Consciousness and Cognition 10: 1-25.

Grondin, S. (2001). From Physical Time to the First and Second Moment's of Psychological Time. Psychological Bulletin, 127:  22-44.

Hanske, G., & Chen, C. (1991). Sex differences in brain organization: Implications for human communication. Inter. J. Neurosci. 28:197-214.

Harrington, D.L., Haaland, K.Y. & Knight, R.T.  (1998). Cortical networks underlying mechanisms of time perception - Journal of Neuroscience, 18: 1085-1095

Hedges, L. & Nowell, A. (1995). Sex differences in mental test scores, variability and numbers of high-scoring individuals. Science 269:41-45.

Jacobs, L.F., Gaulin, S.J., Sherry, D.F. & Hoffman, G.E. (1990). Evolution of spatial cognition: sex-specific patterns of spatial behavior predict hippocampal size. Proc Natl Acad Sci U S A; 87: 6349–6352

Jaffe, K. Urribari, D., Chacon, G.C., Diaz, G., Torres, A., & Herzog, G. (1993). Sex linked strategies of human reproductive behavior. Social Biology 40: 61-73.

Kempel, P., Gohlke,B., Klempau, J., Zinsberger, P., Reuter, M. & Hennig, J. (2005). Second-to-fourth digit length, testosterone and spatial ability. Intelligence, 33, 215-230

Kerns, K.A. & Berenbaum, S.A. (1991). Sex differences in spatial ability in children. Behav. Genetics 21: 383-396.

Kimura, D. (1999). Sex and Cognition. Cambridge, MA, MIT Press, 217 pp.

Kimura, D. (2002). Sex Hormones Influence Human Cognitive Pattern. Neuroendocrinology Letters, 23 sup 4:67–77.

Nagae, S. (1985). Handedness and sex differences in the processing manner of verbal and spatial information. Amer. J. Psychol. 98: 409-420.

Petitot, J. (1989). Perception, cognition and morphological objectivity. Contemporary Music Review, 4: 171 – 180

Poston, T & Stewart, I. (2007). Nonlinear modeling of multistable perception. Behavioral Science 23: 318-334)

Rammsayer, TH. (1997). On the relationship between personality and time estimation. Personality and individual differences 23: 739-744

Rammsayer, T.H. & Lustnauer, S. (1989). Sex differences in time perception. Perceptual and motor skills, 68:195-198.

Rammsayer, T.H. & Rammstedt, B. (2000). Sex-related differences in time estimation: the role of personality. Personality and Individual Differences 29: 301-312

Shaw, P. Greenstein, D. Lerch, J. Clasen, L. Lenroot, R., Gogaty, R., Evans, A. Rappaport, J. & Giedd, J. (2006). Intellectual ability and cortical development in children and adolescents. Nature 440: 676-679.

Seguias, D. (2002). Exploración sociológica sobre manejo de tiempo futuro. Tesis de Licenciatura, Sociología UCAB, Caracas

Thom, R. (1994). Pour une théorie de la morphogénèse. En Les sciences de la forme aujourd'hui, Seuil, París, 1994.

Van der Maas, H. L. & Molenaar, P.C. (1992). Stagewise cognitive development: An application of catastrophe theory. Psychological Review. 99: 395-417

Walsh, V. (2003). Time: the back-door of perception. TRENDS in Cognitive Sciences, 7: 336-338

Xie, Y and Shauman, K.A. 2003. Women in Science Career Processes and Outcomes. Harvard University Press, Cambridge, MA, 336 pp.

Younger, M. & Warrington, M. (2002). Single-sex teaching in a co-educational comprehensive school in England: an evaluation based upon students' performance and classroom interactions. British Educational Research Journal, 28: 353-373.

Zimbardo, P & Boyd, J. (1999). Putting time in perspective: A valid, reliable individual-difference metric. Journal of Personality and Social Psychology 6: 1271-1299.

**Table 1**: Sex differences in time perception and in average school grades. Student's t-tests estimate sex differences. Italian was participants' mother tongue (Ticino is an Italian-speaking canton in Switzerland); English and German were studied as second languages.

|  | Mean for males | Valid N | Mean for females | Valid N | t-value | df | p |
|---|---|---|---|---|---|---|---|
| **Absolute error** | 6.12 | 226 | 6.68 | 267 | -1.39 | 491 | 0.1645 |
| **Scalar error** | -2.0 | 226 | -0.36 | 267 | -2.53 | 491 | 0.011 |
| **Grades in:** |  |  |  |  |  |  |  |
| Italian | 4.17 | 223 | 4.38 | 265 | - 4.72 | 486 | 0.000003 |
| English | 4.06 | 225 | 4.26 | 265 | - 3.34 | 488 | 0.0009 |
| German | 4.28 | 225 | 4.42 | 264 | - 2.45 | 487 | 0.0147 |
| Mathematics | 4.10 | 224 | 4.18 | 264 | - 1.06 | 486 | 0.2909 |
| Natural Science | 4.36 | 225 | 4.31 | 265 | 0.94 | 488 | 0.3496 |
| Economics | 4.33 | 226 | 4.36 | 267 | - 0.46 | 491 | 0.6490 |
| Physical Education | 5.06 | 219 | 4.88 | 251 | 4.96 | 468 | 0.000001 |

**Table 2**: Comparison of scalar errors in assessment of the 3 s sound between students in Education, Sociology and International Relations (A) and Accounting, Economy and Engineering (B) in Caracas.

|  | Mean error A | n | Mean error B | n | t-value | p |
|---|---|---|---|---|---|---|
| **Females** | 0.47 | 146 | 0.24 | 62 | 1.06 | 0.288 |
| **Males** | 0.21 | 21 | -0.69 | 57 | 2.95 | 0.004 |

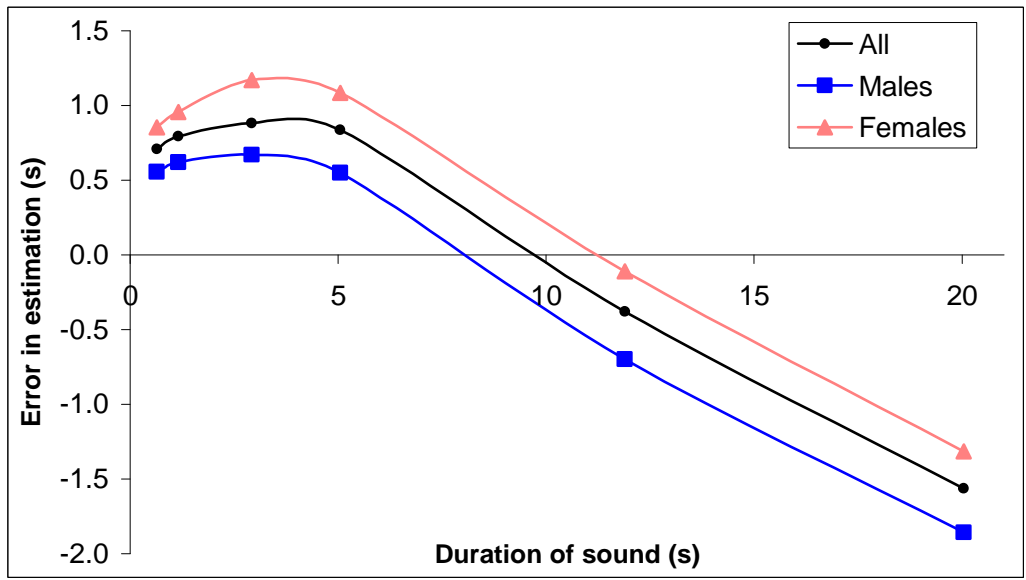

**Figure 1:** Average scalar error (in seconds) for sounds of different duration. Data are from 501 individuals, except for sounds shorter than 5 seconds, which are from a sub-sample of 232 students. Female scores differ significantly from male scores at $p<0.05$ for 0.5 s; $p<0.01$ for 1 and 12 s; and $p<0.001$ for 3 and 5 s.

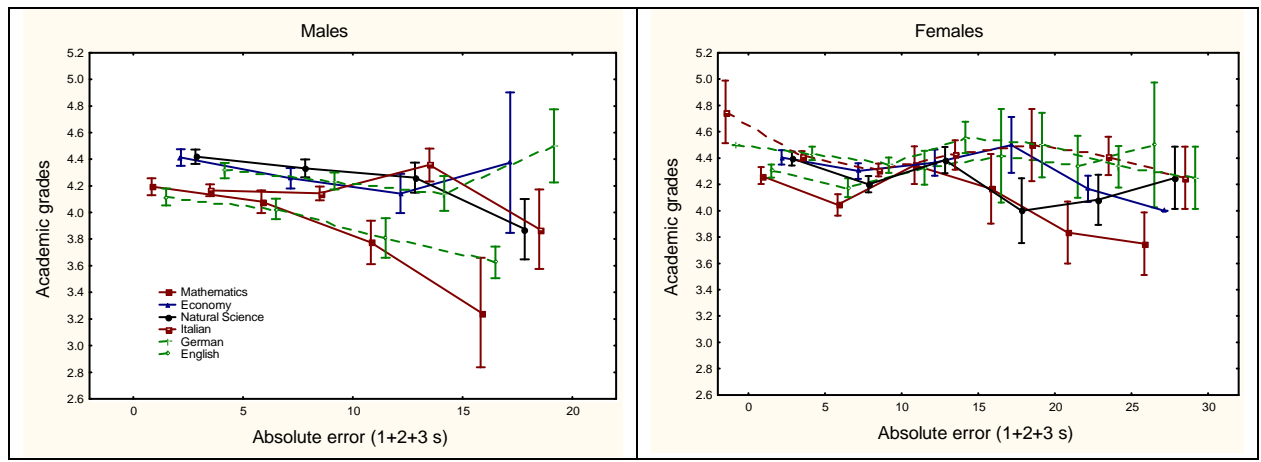

**Figure 2:** Average grades plotted against the sum of absolute errors in time estimates of 1, 2 and 3 s sounds. The error bars represent the standard error of the mean.

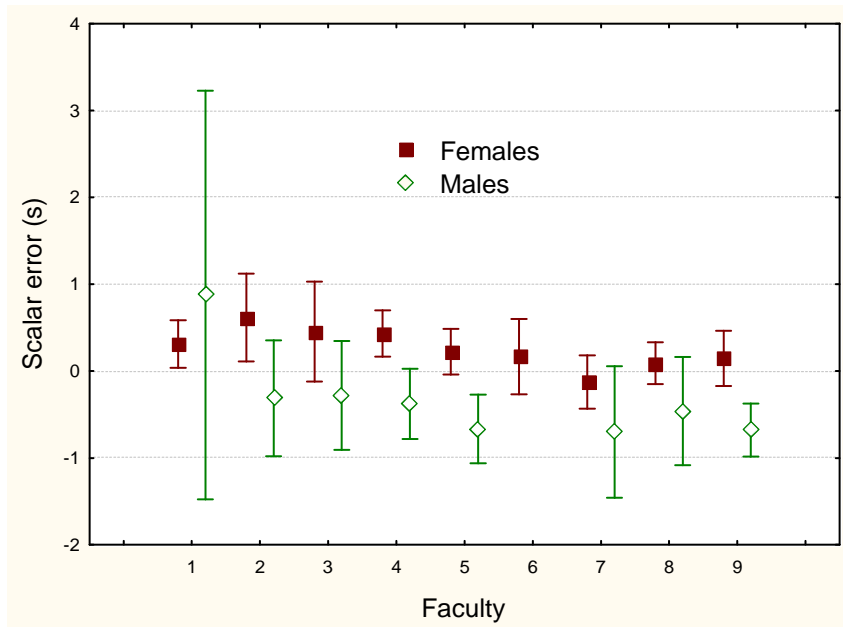

**Figure 3:** Average scalar errors in time estimates for males and females in different faculties. The error bars represent the 95% confidence interval. Faculties are 1: Education, 2: International relations, 3: Philosophy, 4: Sociology, 5: Accounting, 6: Psychology, 7: Economy, 8: Social communication, 9: Engineering.